# minimal effective theory for phonotactic memory: capturing local correlations due to errors in speech

Paul Myles Eugenio

Sept 4 2023

**Abstract**

Spoken language evolves constrained by the economy of speech, which depends on factors such as the structure of the human mouth. This gives rise to local phonetic correlations in spoken words. Here we demonstrate that these local correlations facilitate the learning of spoken words by reducing their information content. We do this by constructing a locally-connected tensor-network model, inspired by similar variational models used for many-body physics, which exploits these local phonetic correlations to facilitate the learning of spoken words. The model is therefore a minimal model of phonetic memory, where "learning to pronounce" and "learning a word" are one and the same. A consequence of which is the learned ability to produce new words which are phonetically reasonable for the target language; as well as providing a hierarchy of the most likely errors that could be produced during the action of speech. We test our model against Latin and Turkish words.

The code is available on GitHub [1].

Consider a game with the following rules: You hear the word "**kitten**", and are asked to produce additional words which start with the same sound "**ki-**". You might produce "**kitchen**", "**kindergarten**", "**kitty-cat**", but you are unlikely to produce "**kipwa**" off the tip of your tongue. While perhaps not surprising, as you have not likely heard the word "**kipwa**" with any frequency (or at all), it is unusual from an informational perspective. If we treat a word as a collection of discrete sounds, then the number of possible combinations of those sounds (i.e words), scales exponentially with the number of sounds in said word. Writing the integer number of sounds a speaker can produce as $d$, and the number of sounds in a word as $N$, then the total number of configurations of said sounds is the number $d^N$. Therefore, taken at random, the word "**kipwa**" (ipa: /kɪpwa/) is exponentially more likely (by a factor $d^{11}/d^5 = d^6$) to be observed than "**kindergarden**" (ipa: /kɪndərgɑrdn/).

Words of real languages are not random configurations however, but instead reflect a history of change governed by a number of factors. This includes language exchange between speaker groups due to migration and trade, along with many other disparate and unpredictable historical events; as well as innovations in a language which may be as random as the coining of a new word, e.g "kipwa". Nevertheless, however arbitrary the origins of spoken words may be, inevitably all words face change governed by the economy of speech.

In order for words to survive in a language, they need be actively communicated, yet no process of communication is error free. However, because of the arbitrary nature of words, if context permits it without producing confusion, it is possible for an error in a spoken word to be produced, without disrupting the communication of the underlying idea that the word represent. For example, "going to" (/ɡoʊɪŋ tu/) → "gonna" (/ɡɑnə/) when used to indicate future actions [2]. In this context, the shortening of the word has the effect of removing redundancy, and therefore has the the added benefit of reducing its overall information content (by $d^{-3}$). It is a compression that involves a change of representation without loss of meaning, and it arises simply due to the need to speak more quickly and use less energy.

Today, "gonna" is a standard feature of the spoken English language, but at its conception, unbiased by any previous use by other English speakers, it could only have been one of a number of possible errors produced in lieu of "going to". Of errors with the same number of sounds as "gonna", are "gopa" (/ɡɑpə/) and "goda" (/ɡɑdə/) and technically even "mdjij"(/mdʒɪdʒ/). But the sound changes (or sequence of sound changes) that gave rise to today's "gonna" were not purely random. Crucially, in order for an instance of error to not hinder communication, the error produced must be *local* – in that it involves a change of sounds within the word, but not a global remaking of the word (i.e not /ɡoʊɪŋ tu/→/mdʒɪdʒ/). Too large a change in the word makes it impossible for the listener to interpret the meaning. Even with the aid of context, such errors may be uncorrectable.

Consider also that the physical constraints of the human vocal apparatus, in addition to a given speaker's trained ability, make certain successive pairs of sounds harder (or less usual) to be produced. This makes certain errors unlikely relative to others. When such constraints are reckoned with locality, a picture emerges of a *local phonetic environment*, in which the interactions between neighboring sounds dictate change. At a basic level, this may involve a change in sounds in order to lower the net actual energy cost of producing speech – e.g the change of /n/ in "in probable" (/ɪnˈprɑbəbəl/) to /m/ in "improbable" (ɪmˈprɑbəbəl). It is simply less physical work to pronounce /m/ before /p/, because both are labial. However, *energy* of interacting sounds need not be purely the physical work cost, but may more abstractly reflect a degree of correlations in the *current* state of the word (i.e given /p/ follows, /m/ is more likely to precede than /n/).

It is the purpose of this manuscript to define the meaning of this energy (see similarly Ref [3]), and therefore construct a statistical model of interacting sounds, analogous to a physical model of strongly interacting local moments.

More importantly, we use this model to demonstrate that local phonetic correlations, arising due to historical language change, can be a means of storing words into memory; and further still reduce the search complexity for retrieving trained words from exponential ($d^N$) to polynomial ($Nd$), by allowing us to play a game akin to the one initially described. Therefore, words can be stored "at the tip of the tongue", allowing a speaker to produce words by searching over those sound configurations which reduce the overall energy set by their local interactions. The interactions which are variationally trained against given words. If we instead consider a listener (as opposed to a speaker), we can reverse this game, which has the meaning of a listener predicting the word before it terminates, i.e using partial information to define conditional probabilities and guess the word.

## 1 The Speaker

To start, it will be useful to consider a language consisting of only three sounds: a vowel ($a$), an unvoiced ($t$) and a voiced ($d$) stop. We write this set of sounds as $\mathcal{S} = \{a, t, d\}$. Sounds are combined into words, which we take here to mean any configuration of sounds. We associate with each sound $s$ in $\mathcal{S}$ a unit vector $n_s$, which is a one-hot vector of dimension $d$ equal to the number of sounds in $\mathcal{S}$. For example here: $n_a = (1,0,0)$, $n_t = (0,1,0)$, and $n_d = (0,0,1)$. A word is constructed as a string of sounds, for example *ata* is represented by the tensor $\boldsymbol{w}_{(ata)} = [n_a, n_t, n_a]$. More generally, any $N$ sound word takes the form $\boldsymbol{w} = [\hat{n}(1), \hat{n}(2), \hat{n}(3), \cdots, \hat{n}(N)]$, where $\hat{n}(x)$ is the placeholder for the sound at position $x$. The index $x$ also represents the progression of time from left to right in a word, reflecting the order in which sounds are communicated during the action of speech.

Let $r$ be an integer distance. For some sound $s$ at $x$ in the word, its correlations with the sound $s'$ at $x+r$ are measured by the quantity $g_{s,s'}(r,x) = n_s^T g(r,x) n_{s'}$, which form the elements of the $d \times d$ correlation matrix $g(r,x)$. We also call this the interaction matrix [3], and henceforth refer also to correlations as interactions. In our 3-sound language, this takes the explicit form

$$g(r,x) = \begin{pmatrix} g_{aa}(r,x) & g_{at}(r,x) & g_{ad}(r,x) \\ g_{ta}(r,x) & g_{tt}(r,x) & g_{td}(r,x) \\ g_{da}(r,x) & g_{dt}(r,x) & g_{dd}(r,x) \end{pmatrix}. \tag{1}$$

which for the purposes of this paper we considered to be strictly positive. This prescribes to each word an energy

$$\mathcal{E}(\boldsymbol{w}) = \sum_x \sum_r \hat{n}(x)^T \big(g_0 - g(r,x)\big) \hat{n}(x+r), \tag{2}$$

where constant $g_0$ is an arbitrary shift in the definition of energy. For the purposes of this discussion, it will suffice to consider the case when $g_0 = 0$. We had argued earlier that an instance of language change due to an error in speech is necessarily local. Since we are considering local correlations which arise due to error in speech, this constrains

$$\overline{g}(1,x) \geq \overline{g}(2,x) \geq \cdots \geq \overline{g}(n-1,x) > \overline{g}(n,x) \gg \cdots, \tag{3}$$

where $\overline{g}(r,x)$ is the average over all the elements of $g(r,x)$. This guarantees that there is some $r > n$ beyond which we can accurately truncate Eqn 2 in $r$. The strongest correlations are between neighboring sounds $g_{s,s'}(1,x)$, followed by next-to-nearest $g_{s,s'}(2,x)$, etc. For the purposes of this paper, we do not consider beyond next-to-next-to nearest neighbor $g_{s,s'}(3,x)$. And further, even though there is no constraint to do so, we will consider purely uniform interactions (i.e constant in $x$: $g_{s,s'}(r,1) = g_{s,s'}(r,2) = \cdots$). (We will return later to non-uniform interactions.)

Within these constraints, there remains a significant freedom in the values of the elements of $g(r)$. One could implement further constraints, such as for our example language, where we may choose $g_{VV}(2) > g_{VV}(1) > g_{CC}(1)$ and $g_{CV}(1) \simeq g_{VC}(1) > g_{CC}(1)$, where $C$ is a consonant ($t/d$) and $V$ is a vowel (here only $a$). This has the effect of guaranteeing a predominantly $CV$-type syllable structure for our language, by making words constructed with that syllable pattern the lowest in energy, e.g $\mathcal{E}(\boldsymbol{w}_{(tata)}) < \mathcal{E}(\boldsymbol{w}_{(atta)})$. Or similarly, $g_{ad}(1) > g_{at}(1)$ may manifest as a voicing of stops between vowels, e.g $\mathcal{E}(\boldsymbol{w}_{(ada)}) < \mathcal{E}(\boldsymbol{w}_{(ata)})$.

More generally, we should think of the elements $g_{s,s'}(r)$ as spanning a variational manifold, whose measure is the energy Eqn 2. This is to say that the relative energies describe an ordering of all $\boldsymbol{w}$'s; where the lowest energy (a.k.a ground) words are most likely to be observed, followed by low-lying words, then words at higher energy. The exact statistical distribution of the words is not as critical as the relative ordering of their probabilities. For example, one could choose to use a thermal distribution [3], i.e the probability of observing $\boldsymbol{w}$ with the form

$$\mathcal{P}(\boldsymbol{w}) = \frac{1}{\mathcal{Z}} \exp\left(-\beta \mathcal{E}[\boldsymbol{w}]\right), \tag{4}$$

where $\mathcal{Z}$ normalizes the distribution to 1 (or 100%), and $0 < \beta < \infty$ is a free parameter that changes the relative probabilities without changing their order. (Taking $\beta \to \infty$ guarantees the ground state.)

In practice, we do not constrain the elements of $g(r)$ by hand, but instead train our model by minimizing the energy of a set of input words. The resulting $g_{s,s'}(r)$ encode information about the patterns in these input words, and if the sample is sufficiently large, allows it to learn the phonetic rules of the language. It is the purpose of this paper to show that the structure of natural spoken languages is such that the critical phonetic rules can be learned given only a handful of input words. And that further, new words can be constructed which satisfy these local phonotactic constraints, some of which are real words of the target language against which the $g_{s,s'}(r)$ have *not* been trained. We call this model (Eqn 2) and its collection of constraints, learned from input words, as the *speaker*.

## 2 Words and Pseudowords

Most importantly, we find that the $g_{s,s'}(r)$ encodes not only the phonetic rules of the language, but can also be used to regenerate the input words, and thus act as a type of phonetic memory. The key to this trick lies in how the $g_{s,s'}(r)$ define conditional probabilities. This is done with the aid of an initial sound, e.g /kɪ/, which acts as a (left) boundary condition on Eqn 2,

$$\begin{aligned}\mathcal{E}_{\text{kɪ}}(\boldsymbol{w}) &= n_{\text{k}}(1)^T\big(g_0 - g(1)\big)n_{\text{ɪ}}(2) + n_{\text{k}}(1)^T \sum_{r>1}\big(g_0 - g(r)\big)\hat{n}(1+r) \\ &+ n_{\text{ɪ}}(2)^T \sum_{r>0}\big(g_0 - g(r)\big)\hat{n}(2+r) + \sum_{x>2}\sum_{r>0}\hat{n}(x)^T\big(g_0 - g(r)\big)\hat{n}(x+r).\end{aligned} \quad (5)$$

This is equivalent to $\mathcal{E}_{\text{kɪ}}(\boldsymbol{w}) = \mathcal{E}([n_{\text{k}}, n_{\text{ɪ}}, \boldsymbol{w}])$, where $\boldsymbol{w}$ is the placeholder for the list of remaining unset sounds. In this example, the second and third terms correspond to the energy due to the interaction between the boundary and the sounds which follow. Within the physically analogy, they can be understood as generically-non-uniform one-body potentials, which arise within the effective theory for the remaining sounds. These terms define a new space of words, organized by their energies, under the constraint that the initial sounds be those fixed. If the speaker has been sufficiently trained against a set of input words, containing words which begin with /kɪ/, then those words exists as low-energy solutions within the various $N$-sound problems defined by Eqn 5.

In order to search this space of low-energy words, we need only grow the word sound-by-sound, under the constraint that the next sound following /kɪ/ minimizes Eqn 5. As a demonstration of what this might look like in practice, consider a speaker trained against the set of input words: {kɪndərgɑrdn,kɪd}. We then search for the lowest energy next sound of Eqn 5, of which there are $d$ possibilities. Given the input words, we can expect the two lowest energy sounds to be either /d/ or /n/, for /kɪd/ or /kɪn/ respectively. Which of the two is the true ground state depends on the specifics of the training, which for the purposes of this demonstration, we take to produce $\mathcal{E}_{\text{kɪ}}([n_{\text{n}}]) < \mathcal{E}_{\text{kɪ}}([n_{\text{d}}])$, such that the speaker constructs /kɪn/. Continuing sound-by-sound looks like /kɪ/→/kɪn/→/kɪnd/→ ··· →/kɪndərgɑrdn/, where the speaker eventually produces the input word.

We can then search for the low-lying (but non-ground) word /kɪd/ by introducing a term into Eqn 5 which penalizes the ground word, here being partial /kɪn/, for example,

$$\delta\mathcal{E} = \Big(\lambda|\boldsymbol{w} - \boldsymbol{w}_{\text{/kɪn/}}|^2 + \lambda^{-1}\Big)^{-1}, \quad (6)$$

where $\lambda \to \infty$. More generally, this trick allows us to search for all input words, regardless of their initial sound. This is not necessary if our training words are only those two described, then we could have started with /k/, where /ɪ/ would be guaranteed to be the ground state for the proceeding sound, giving /kɪ/. If instead our set of input words contains words which do not begin with /kɪ/, but (say) also one which begins with /kɑ/, then we can penalize /kɑ/ in order to search over those words which begin with /kɪ/. Therefore by searching over the $d$ possible sounds $N$ times, the speaker need only search over an order $\mathcal{O}(Nd)$ space of words which are stored "at the tip of the tongue".

It might be surprising that a speaker playing this game might reproduce /kɪndərgɑrdn/, a longer word than /kɪd/, even though the latter has fewer sounds. If we use the values of the energies to define a conditional probability given /kɪ/, i.e $\mathcal{P}_{\text{kɪ}}(\boldsymbol{w})$, then this example implies $\mathcal{P}_{\text{kɪ}}([n_{\text{n}}]) > \mathcal{P}_{\text{kɪ}}([n_{\text{d}}])$. In other words, local correlations may make longer input words more likely to be reproduced than shorter words. This works because only the relative energies within the same conditional sector (i.e with same boundary sounds) matter for this game. The total probability of observing /kɪndərgɑrdn/ given /kɪ/ is the product

$$\mathcal{P}_{\text{kɪ}}([n_{\text{n}}, n_{\text{d}}, n_{\text{ə}}, n_{\text{r}}, n_{\text{g}}, n_{\text{ɑ}}, n_{\text{r}}, n_{\text{d}}, n_{\text{n}}])$$
$$= \mathcal{P}_{\text{kɪ}}([n_{\text{n}}])\mathcal{P}_{\text{kɪn}}([n_{\text{d}}])\mathcal{P}_{\text{kɪnd}}([n_{\text{ə}}])\mathcal{P}_{\text{kɪndə}}([n_{\text{r}}])\mathcal{P}_{\text{kɪndər}}([n_{\text{g}}])$$
$$\times \mathcal{P}_{\text{kɪndərg}}([n_{\text{ɑ}}])\mathcal{P}_{\text{kɪndərgɑ}}([n_{\text{r}}])\mathcal{P}_{\text{kɪndərgɑr}}([n_{\text{d}}])\mathcal{P}_{\text{kɪndərgɑrd}}([n_{\text{n}}]). \qquad (7)$$

This well-established sequence for word (and even sentence) generation, proposed first by Ref [4], is here a natural consequence of the local interactions. The process of training, by minimizing the energy of the input words, guarantees that the conditional probabilities for observing the intermediate states (e.g /kɪndərgɑ/ given /kɪndərg/) are close to unity.

Put more simply, the model needs to learn to pronounce partial words in order to reproduce complete words. This has the consequence that the speaker necessarily learns words which are not in the training set. This ability extends beyond the necessary reproduction of partial input words, but also to novel words of any length, which stem from any boundary, and which (at low energies) are phonetically correct for the target language. (Note that for the speaker, the target language is defined entirely by the input words.) This feature of generality owes itself to the uniformity of the interactions in Eqn 2. Rather than rote memorization of sounds, the speaker learns words by only learning the local correlations which define its phonotactic constraints. The process of "learning a word" and "learning to pronounce" are one in the same [5].

This behaviour is demonstrated in the relationship between Fig's 1a-1c and 2a-2b for a speaker trained against a selection of Latin nouns [6]. In Fig's 2a-2b we see the branch-like conditional spaces generated by playing the game laid out in this section. By learning correlations which are statistically significant for the input words, the speaker is able to reproduce the input words, while additionally learning variations in those input words. We call all learned words which are not input words *pseudowords*[1], which includes such variations in the input words. All pseudowords can be understood by their spatial relationship to trained input words in the branching space. Ground words, which are strongly biased to be input words (or partial input words), are at the bottom, with no words below it in energy. As previously explained, these are the most likely words to be produced given the boundary. The word above them in energy is either themself an input word (such as /kɪd/ being above /kɪn/ in our example), or is the most probable error given the phonotactic constraints.

---
[1] The more commonly used definition of *pseudoword* [7] is closer to what we refer to here as a low-energy pseudoword.

Not all the pseudowords are meaningful, as the speaker learns words phonetically without any sense of grammar or meaning; however, some actual Latin words (such as the feminized "serva"/"servā") not only appear at low energies, but additionally the speaker intuits some correct morphological forms (e.g "servās, servārum"). In Latin "servārum" is the plural possessive of "servā", meaning "female servant/slave". Mathematically, the ending "rum" is one of the $21 \times 21 \times 21 = 9261$ combinations of 3 sounds which could follow "servā", and $194,481$ possibilities following the root "serv". The speaker learns the root itself as the partial construction of input words like "servus", and its morphology is intuited by analogy with "īnsula"/"īnsulārum". But the fact that "servārum" and "īnsulārum" have parallel morphologies should be understood as a feature of the language. Where such forms evolved from local errors in a similar process to "going to"→"gonna", or are analogies learned by speakers, who carried them over from other words where such an error occurred [2].

## 3 Gibberish

Too my knowledge, all languages have some degree of morphology. In English, this occurs in words like "happy"/"happily"/"happiness" and in the subject-object relationship of common pronouns like "he"/"him" or "they/them". By comparison to English, Turkish has an extreme degree of morphology, such that the equivalent of an entire English sentence can be expressed in a single word. For example [2], "şehirlileştiremediklerimizdensiniz" meaning "you are one of those whom we can't turn into a town-dweller". Technically, this process can be extended to indefinitely-long Turkish words, but such construction tend to become non-sensical. A similar phenomenon happens in English with made-up constructions like "fartological" ("fart-ology-ical"), roughly meaning "valid within the study of farts"; which can be extended in a seemingly-meaningful yet meaningless way as "fartologicallinessing" ("fartological-ly-ness-ing").

A definition of "words" which links sound to meaning cannot account for such constructions. While the ultimate use of words by humans is to express meaning, the ability to construct phonetically reasonable strings of gibberish is evidence for a system of the brain which defines words devoid of meaning. Phonetic memory of the type employed here is just such a mechanism, where (again) the process of "learning a word" and "learning to pronounce" are one in the same. By learning the phonotactic constraints, new words can be constructed by tying together learned chunks of other words in a phonetically reasonable way[2]. This is crucial for a language with a high degree of morphology, like Turkish.

---

[2]See Ref's [7, 5, 8, 9] for a review of discrete phonological models at different scales: sounds, morphemes, words.

Consider Fig 3a for a speaker trained against a collection of Turkish nouns [10, 11, 12]. The input word "güzel" (translated "beautiful"/"delicious") lies at the ground-level of the branching space that starts with "g". But because there are no input words which start with "güzel" as its base, asking the speaking to grow the word beyond "güzel" necessarily forces it to produce a pseudoword. Because the speaker constructs the lowest energy pseudoword given the boundary, continuing the game sound-by-sound generates either: (1, by accident) a meaningful construction or (2) a meaningless one which sounds as reasonable as possible. For the speaker in Fig 3a, this happens to be the prior: "güzeler", an approximate to Turkish "güzeller", meaning "beautifuls"/"beauties".

While only an approximation of the actual Turkish plural "güzeller", we can confidently see it as a pluralization because it learned the plural marker "-ler" by anology with input words which do contain it [10]. (Note that more training words are needed to learn the "l/ll" distinction.) The plural markers "-ler/-łar" are learned correlations; which in the physical analogy means that the interactions between their sounds push those configurations down in energy. Of the two forms of the plural markers "-ler/-łar", the prior "güzeler" wins out. This is because of the vowel harmony present in Turkish, which as a rule requires that the vowels of attached word-endings necessarily match the type of its preceding stem: front (i/e/ö/ü) or back (ı/a/o/u). In other words, Turkish speakers (like speakers of all languages) construct words which are easy to pronounce. What is *easy/hard* may depend on the physical structure of the mouth, but can also reflect the degree of usualness learned through exposure and practice.

As we further probe the speaker for the lowest-energy next sounds, it continues to generate recognizable morphemes in a way which satisfies both the Turkish syllable structure and vowel harmony, producing "güzeler-sünüz-elim". Eventually our speaker produces a collection of sounds "eerez" which isn't a recognizable morpheme, and which has a VV-type syllable pattern not typical of Turkish. However, in Fig 3a, we plot the local energy due to the interactions – in units labelled by triangles, with no triangle implying zero energy – and find a jump in energy where this occurs. Thus, the speaker knows that this is an unusual construction, it just doesn't know of any better sounds to produce. In effect, it's stuck as there are no easier to pronounce sounds which fit the phonetic environment following "-elim". Thus it may be better to interpret the gibberish as breaking up into multiple words:
"güzelersünüzelime erezelime erez", to reflect the absence of correlation between those parts of the string. (This also lowers the energy.) Ultimately the pattern repeats due to the lack of correlations, and the short-ranged nature of the interactions. We call this the *steady state of gibberish.*

One way to break the steady-state and add variety into the gibberish is to play the next-sound game, but asking for a next-to-lowest (instead of a lowest) sound when growing the word. This forces the speaker up higher in the branching space, similar to the process of searching for the input word /kɪd/ by penalizing /kɪn/ in our previous example. Thus allowing for a greater variety of gibberish, by demanding that the speaker explore more low-lying (but not necessarily ground) pseudowords in the branching space. We demonstrate this in Fig 3b, starting with the root "kɪ", and rolling a 5-sided die to determine vertical motion in the branching space: 1/5 grows by next-to-lowest sound & 4/5 grows the lowest energy sound. We find that the speaker proceeds through the gibberish in a phonetically reasonable way, producing bunches of sounds which are recognizable morphemes learned from the input words.

Notice that repeating this process with the farthest – then next farthest, etc – interactions set to zero demonstrates how the morphemes are encoded by the speaker. This is most pronounced upon turning off the next-to(-next-to)-nearest interactions ($g(3) = g(2) = 0$), which damages the speaker's ability to produce consistent vowel harmony. Turning off all interactions ($g(r) = 0$ for all $r$) erases all the correlations, by making all sounds have the same energy. (Note in Fig 3b this is seen in repeating "e/i", which is the chosen default sound ordering, which manifests when both sounds are at equal energy. The occurrence of either "e" or "i" reflects the 1/5 statistics of the dice roll.)

## 4   Conclusion

In this manuscript, we demonstrated that local interactions are sufficient for organizing phonetic information in a dictionary-like fashion, code for which has been made publicly available to facilitate the reader's further exploration [1]. Our principal argument is that the local nature of error correction, along with constraints on the production of neighboring sounds, gives rise to words which are locally organized. This significantly reduces the information volume of words, thus allowing speakers to organize information in a productive/predictive fashion useful to speaking/listening. It is a statement that languages evolve in a way which is advantageous to speakers, such that a linear locally-connected model can act as memory.

More complicated models using non-linear variational ansatz, such as neural networks, could likewise capture this phenomenon [3, 13, 14, 8]; in fact, one interpretation of Eqn 2 is that it is analogous to the synapse of a single-layer Boltzmann machine [14, 3], but where the non-linear nodes have been stripped away. Perhaps more appropriately, it is classical analog of a matrix product operator [15, 16, 17, 18, 19, 20, 21], which were developed to solve a class of exponentially-hard quantum many-body physics problems. The next-sound game employed here is similar to the density matrix renormalization group [16, 15], where contracting all but a local pair of matrices produces an effective energy (like Eqn 5). Upon recognizing that sounds are correlated locally, we immediately get a perfect analogy with a model of a 1D locally-interacting moments (i.e spins), and all the tools developed to study those physics problems immediately apply.

One particular advantage to an energetic approach to phonotactics, is that it provides a measure of distance between sounds in a word. A high energy for a local arrangement of neighboring sounds indicates a lack of local correlation between those sounds. We demonstrated this in Sec 3 by forcing the speaker to continue to produce sounds trailing "güzel", and argued that the hard-to-pronounce (i.e high energy) regions may be best interpreted as a breaking up of the string into multiple smaller strings, "güzelersünüzelime erezelime erez". Within the language of Eqn 2, this is the same as lowering the energy of the string by inserting voids which effectively separate those sounds, thus lowering their local interactions. In this interpretation, local correlations define a sense of "closeness" which may help to distinguish a "word" from other words produced during speech[3].

No doubt, for a speaker stringing together neighboring sounds to produce a word, there needs to be some mechanism to indicate when this procedure should be stopped. Local correlations between sounds are not enough to indicate meaning or grammatical context. Thus this motivates the existence of an external mechanism for searching the conditional spaces, and recognizing the word once it has been produced. All that is required is to know the rough number of sounds, and approximate information about where the boundaries should be placed in order to satisfy the search[4].

A theory of phonetic memory is inherently a theory of language change, as the most likely errors are local variations in learned input words (a.k.a low-energy pseudowords). This also extends to what linguists call the process of "analogy" [2]: While our principal argument is hinged on words evolving due to local errors, in practice this does not happen for all words. Instead this change happens for a subset of popular words, then over time, the errors in these words are reinterpreted by new generations of speakers to be standard [2]. These new generations then carry over those changes to the remaining words by phonetic analogy, thus leading to (for example) the different declensions in Latin (e.g the declension of "īnsula/īnsulārum" & "serva/servārum"). Analogy is a natural consequence of phonetic memory. This is because (1) in order to learn input words, partial input words, which includes morphemes, need to be pushed down in energy; and (2) the interaction between the end of a root and its proceeding morphemes constrains the morphology of that root [5, 22]. More generally, morphology exists in a spectrum of possible errors to a given word. Even if a speaker has never explicitly observed that morphological form, they intuit its shape based on the phonotactic constraints [5]. If in the future they do explicitly observe that form, it is easy to learn, being already low in energy.

There is an argument to be made that one cannot have a theory of the mind that makes no reference to its specific physiology. Building up a biological theory of the mind from the ground up is an ultimate goal; however, as a lesson learned from physics, phenomenological models can be powerful aides in the study of microscopic systems. (This being true even when the microscopic system is completely unknown, such as with the Standard Model of Particle Physics.) A likely candidate for a microscopic theory are the asynchronous neurons of the type proposed by Hopfield [23, 24, 13]. Such neurons are analogous to all-to-all coupled Ising spins, which can encode memories in their interactions, similar to Eqn 2. Because the minimal model Eqn 2 requires only local interactions between moments with spin $d$, we question if it can arise out of the collective interaction of such neurons. In this regard, we consider Eqn 2 to be an effective theory constrained by the informational conditions evident in natural spoken languages.

---

[3]For instance /ɢoʊɪŋ tu/ taking the interpretation of two words because /ŋt/ is unusual pair for an English speaker; as opposed to /ɢɑnə/, which is constructed of more typically syllables.

[4]This may be evident in tip-of-tongue circumstances where one struggles to recall a word like "lobbying" (/ɬɑbiiŋ/), yet knows the word sounds like "loitering" (/ɬɔɪtsɹɪŋ/). Both words are roughly the same length and contain /ɬ/ and /ŋ/ at the start and end of the word.

## acknowledgements

Special thanks to Ceren Burçak Dağ for comments & discussion.

## References

[1] URL: https://github.com/pboxinator/DiscretePhonotactics.

[2] Guy Deutscher. *The Unfolding of Language*. Especially the discussions on analogy; and also the historical development of "gonna" from "going to". Holt Paperbacks, 2005. ISBN: 978-0-8050-8012-4.

[3] Huteng Dai, Connor Mayer, and Richard Futrell. "Rethinking representations: A log-bilinear model of phonotactics". In: *Proceedings of the Society for Computation in Linguistics (SCiL), pages 259-268* (2023). DOI: 10.7275/ebv1-5g73.

[4] E. Colin Che, Morris Hall, and Roman Jakobson. "Toward the logical description of languages in their phonemic aspect". In: *Language, Vol. 29, No. 1 (Jan. - Mar., 1953), pp. 34-46* (1953). URL: http://www.jstor.org/stable/410451.

[5] Tiago Pimentel, Brian Roark, and Ryan Cotterell. "Phonotactic Complexity and Its Trade-offs". In: *Transactions of the Association for Computational Linguistics* 8 (Jan. 2020), pp. 1–18. ISSN: 2307-387X. DOI: 10.1162/tacl_a_00296. eprint: https://direct.mit.edu/tacl/article-pdf/doi/10.1162/tacl\_a\_00296/1923388/tacl\_a\_00296.pdf. URL: https://doi.org/10.1162/tacl%5C_a%5C_00296.

[6] **input Latin words:** īnsula,īnsulam,īnsulae,īnsulā,īnsulās,īnsulārum,īnsulīs, servus,servum,servī,servō,servōs,servōrum,servīs,verbum,verbī,verbō,verba, verbōrum,verbīs,pāstor,pāstōrem,pāstōris,pāstōrī,pāstōre,pāstōrēs,pāstōrum, pāstōribus,ovis,ovem,ovī,ove,ovēs,ovium,ovibus.

[7] David Kemmerer. *Cognitive Neuroscience of Language*. Psychology Press, 2015. ISBN: 978-1-84872-621-5.

[8] P. Boersma, K. Chládková, and T. Benders. "Phonological features emerge substance-freely from the phonetics and the morphology". In: *Canadian Journal of Linguistics* (2022). DOI: 10.1017/cnj:2022.39.

[9] Stephen Merity, Nitish Shirish Keskar, and Richard Socher. *An Analysis of Neural Language Modeling at Multiple Scales*. 2018. arXiv: 1803.08240 [cs.CL].

[10] **input Turkish words:** o,onu,ondan,onłarı,onłara,onłardan,bunu,buna,bundan, bunłarı,bunłara,bunłardan,şunu,şunłarı,şunłara,şunłardan,kimi,kime,kimden, kimlere,kimlerden,bugünü,bugüne,bugünler,kızım,kadınım,güzelim,güzelsin, güzeliz,güzelsiniz,şoförüm,şoförsün,şoförüz,şoförsünüz,doktorum,doktorsun, doktoruz,doktorsunuz,yalnızım,yalnıssın,yalnızız,yalnıssınız. ipa: c = /dʒ/, ç = /tʃ/, ı = /ɯ/, j = /ʒ/, ö = /œ/, r = /ɾ/, ş = /ʃ/, ü = /y/, y = /j/, and we further distinguish between /l/ & /ɫ/ but not /e/ & /ɛ/.

[11] Lewis V. Thomas and Norman Itzkowitz. *Elementary Turkish*. Dover Publications, 1986. ISBN: 978-0-486-25064-9.

[12] Korkut Buğday and (transl.) Jerold Frakes. *Literary Ottoman*. Routledge, 2009. ISBN: 978-0-415-49438-0.

[13] P. Boersma, Benders T., and Seinhorst K. "Neural network models for phonology and phonetics". In: *Journal of Language Modelling* (2020). DOI: 10.15398/JLMV8I1.224.

[14] Paul Boersma. "Simulated distributional learning in deep Boltzmann machines leads to the emergence of discrete categories". In: *Proceedings of the 19th International Congress of Phonetic Sciences* (2019).

[15] Ulrich Schollwöck. "The density-matrix renormalization group in the age of matrix product states". In: (2011). URL: https://doi.org/10.1016/j.aop.2010.09.012.

[16] Steven R. White. "Density-matrix algorithms for quantum renormalization groups". In: *Phys. Rev. B* (1993). DOI: 10.1103/PhysRevB.48.10345. URL: https://link.aps.org/doi/10.1103/PhysRevB.48.10345.

[17] E. Miles Stoudenmire. "A Quantum-Inspired Algorithm for Solving Differential Equations". In: *Journal Club for Condensed Matter Physics* (2022). DOI: 10.36471/JCCM_November_2022_03.

[18] E. M. Stoudenmire and David J. Schwab. "Supervised Learning with Tensor Networks". In: *30th Conference on Neural Information Processing Systems (NIPS 2016), Barcelona, Spain* (). URL: https://proceedings.neurips.cc/paper/2016/file/5314b9674c86e3f9d1ba25ef9bb32895-Paper.pdf.

[19] E Miles Stoudenmire. "Learning relevant features of data with multi-scale tensor networks". In: *Quantum Sci. Technol. 3 034003* (2018). DOI: 10.1088/2058-9565/aaba1a.

[20] Tai-Danae Bradley et al. "Modeling sequences with quantum states: a look under the hood". In: *Mach. Learn.: Sci. Technol. 1 035008* (2020). DOI: 10.1088/2632-2153/ab8731.

[21] J A Reyes and E M Stoudenmire. "Multi-scale tensor network architecture for machine learning". In: *Mach. Learn.: Sci. Technol. 2 035036* (2021). DOI: 10.1088/2632-2153/abffe8.

[22] Paul Boersma and Jan-Willem van Leussen. "Efficient evaluation and learning in multi-level parallel constraint grammars". In: *Linguistic Inquiry Vol. 48, No. 3 (published Summer 2017), pp. 349-388 (40 pages)* (2016). URL: https://www.jstor.org/stable/26799876.

[23] J J Hopfield. "Neural networks and physical systems with emergent collective computational abilities." In: *Proceedings of the National Academy of Sciences* 79.8 (1982), pp. 2554–2558. DOI: 10.1073/pnas.79.8.2554. eprint: https://www.pnas.org/doi/pdf/10.1073/pnas.79.8.2554. URL: https://www.pnas.org/doi/abs/10.1073/pnas.79.8.2554.

[24] Marco Benedetti et al. *Eigenvector Dreaming*. 2023. arXiv: 2308.13445 [cond-mat.dis-nn].

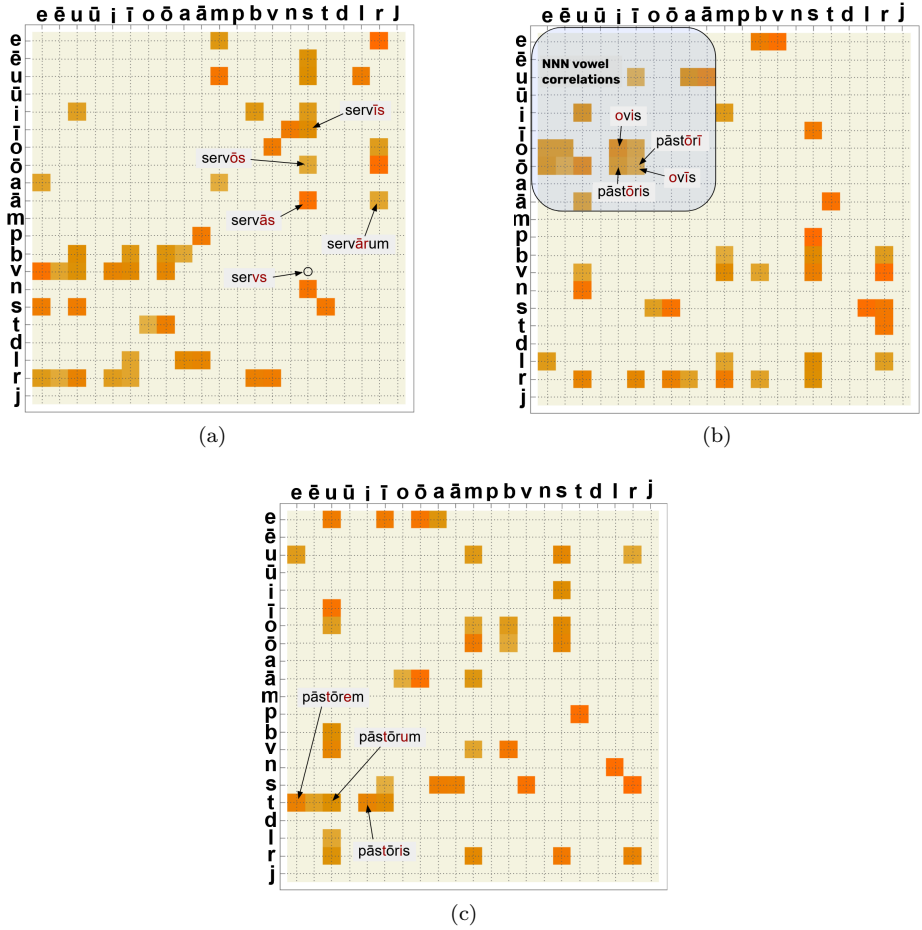

Figure 1: Correlation peaks plotted as $(g_0 - g(r))^{-1}$ for a speaker (Eqn 2) trained against Latin input words [6]: (a) nearest-neighbor $g(1)$, (b) next-to-nearest $g(2)$, and next-to-next-to-nearest neighbor $g(3)$ interactions. Darker color implies stronger correlations. Note the lack of nearest-neighbor vowel-vowel correlations absent in the input words. Additionally note that some short-long vowel pairs (e.g o/ō) exhibit a degree of symmetry – i.e the speaker sees these sounds as similar from context. Each $g(r)$ was trained together using the *gflow_uniform_2* function for 10000 timesteps.

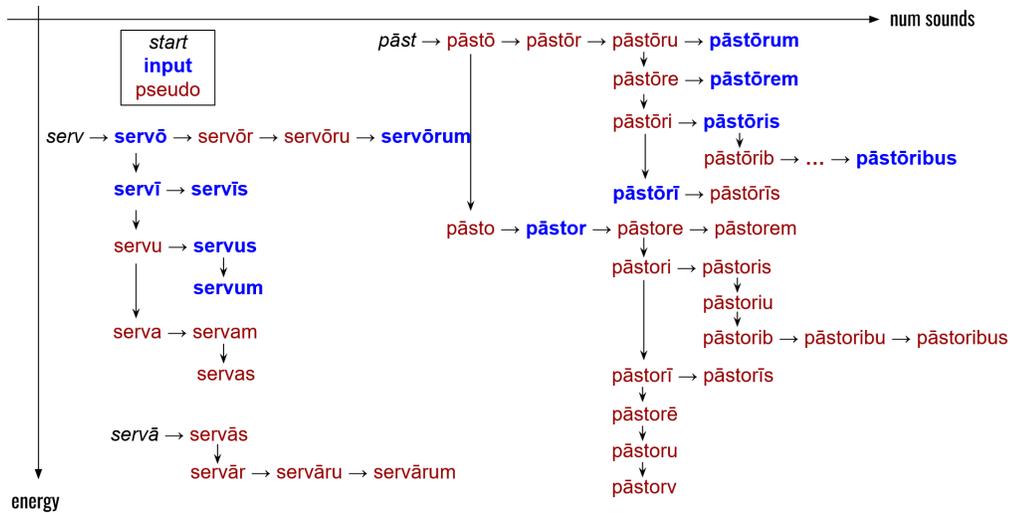

(a)

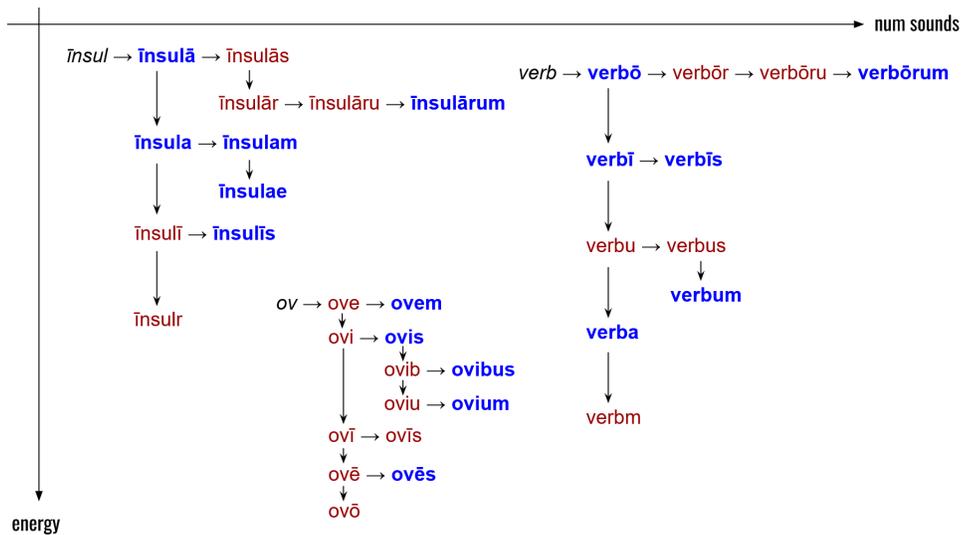

(b)

Figure 2: Branching spaces generated by speaker of Fig's 1a-1c. Words are organized in memory by either growing the word (right arrows) or finding the next most-likely word with the same number of sounds (down arrows). A pseudoword is defined as any word that isn't an input word. Note that low-energy pseudowords tend to be local variations in either the input words or partial input words.

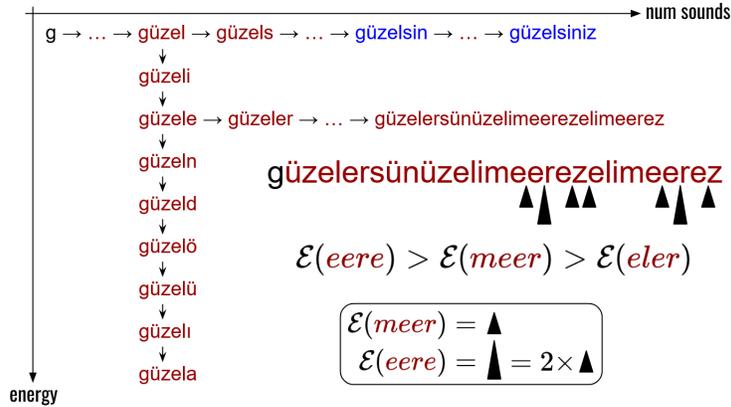

(a)

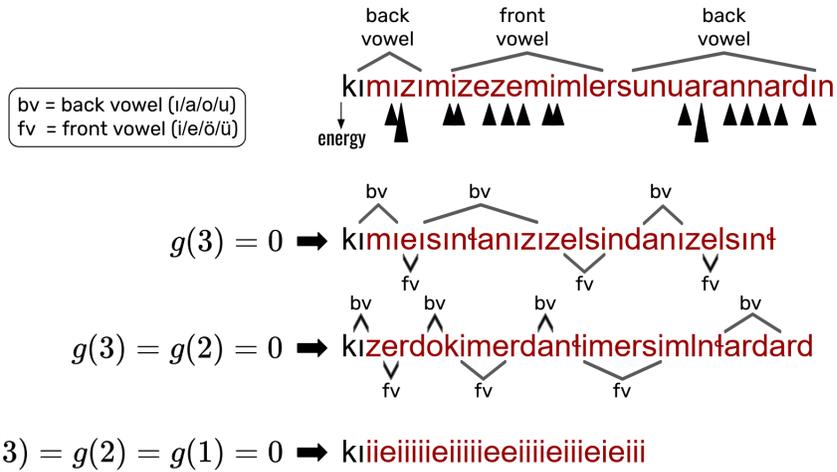

(b)

Figure 3: (a) Speaker trained against Turkish input words [10]. Asking the speaker to reproduce the input word "**güzel**" ("beautiful") and "**güzelsiniz**" ("you are beautiful"); but the speaker learns "**güzeler**" ("beauties") by analogy with words which contain "**-ler**". Continuing to grow the word beyond "**güzeler**" generates a non-sensical but phonetically-reasonable construction (a.k.a gibberish). Local energies defined by Eqn 2 are shown as black triangles which point to the center of interaction. Units are chosen such that one small black triangle equals the default (i.e before training) value of $g_0 = 1$. No triangle implies zero energy; with each small triangle being approximately equal to 1, out of 6 total. (b) Gibberish generated by growing the lowest-energy word (4/5 of the time) or growing the next-lowest-energy word (1/5). Randomness allows for the speaker to explore more low-energy configurations without hitting a steady-state. Low-energy regions (no triangles) tend to be learned morphemes. Turning off the interactions range-by-range demonstates how the local correlations are encoded by Eqn 2. Nearest-neighbor interactions are sufficient to capture the CV and CCV type syllable structure, as well as learn some morphemes (e.g -łar/-ler). But next-to(-next-to)-nearest neighbor interactions are necessary in order for the speaker to consistently produce vowel harmony. Training function was *gflow_uniform_2* for 10000 timesteps.